\documentstyle[twocolumn,aps,epsfig,floats]{revtex}

\begin{document}

\def\d{{\rm d}}
\def\e{{\rm e}}
\def\O{{\rm O}}
\def\half{\mbox{$\frac12$}}
\def\eref#1{(\protect\ref{#1})}
\def\etal{{\it{}et~al.}}

\draft
\tolerance = 10000

\setcounter{topnumber}{1}
\renewcommand{\topfraction}{0.9}
\renewcommand{\textfraction}{0.1}
\renewcommand{\floatpagefraction}{0.9}

%Fixing abstract in twocolumn mode
\twocolumn[\hsize\textwidth\columnwidth\hsize\csname @twocolumnfalse\endcsname

\title{Scaling and percolation in the small-world network model}
\author{M. E. J. Newman and D. J. Watts}
\address{Santa Fe Institute, 1399 Hyde Park Road, Santa Fe, NM 87501}
\date{May 6, 1999}
\maketitle

\begin{abstract}
  In this paper we study the small-world network model of Watts and
  Strogatz, which mimics some aspects of the structure of networks of
  social interactions.  We argue that there is one non-trivial length-scale
  in the model, analogous to the correlation length in other systems, which
  is well-defined in the limit of infinite system size and which diverges
  continuously as the randomness in the network tends to zero, giving a
  normal critical point in this limit.  This length-scale governs the
  cross-over from large- to small-world behavior in the model, as well as
  the number of vertices in a neighborhood of given radius on the network.
  We derive the value of the single critical exponent controlling behavior
  in the critical region and the finite size scaling form for the average
  vertex--vertex distance on the network, and, using series expansion and
  Pad\'e approximants, find an approximate analytic form for the scaling
  function.  We calculate the effective dimension of small-world graphs and
  show that this dimension varies as a function of the length-scale on
  which it is measured, in a manner reminiscent of multifractals.  We also
  study the problem of site percolation on small-world networks as a simple
  model of disease propagation, and derive an approximate expression for
  the percolation probability at which a giant component of connected
  vertices first forms (in epidemiological terms, the point at which an
  epidemic occurs).  The typical cluster radius satisfies the expected
  finite size scaling form with a cluster size exponent close to that for a
  random graph.  All our analytic results are confirmed by extensive
  numerical simulations of the model.
\end{abstract}

\vspace{1cm}

%Fixing abstract in twocolumn mode
]

\section{Introduction}
Networks of social interactions between individuals, groups, or
organizations have some unusual topological properties which set them apart
from most of the networks with which physics deals.  They appear to display
simultaneously properties typical both of regular lattices and of random
graphs.  For instance, social networks have well-defined locales in the
sense that if individual~A knows individual~B and individual~B knows
individual~C, then it is likely that A also knows C---much more likely than
if we were to pick two individuals at random from the population and ask
whether they are acquainted.  In this respect social networks are similar
to regular lattices, which also have well-defined locales, but very
different from random graphs, in which the probability of connection is the
same for any pair of vertices on the graph.  On the other hand, it is
widely believed that one can get from almost any member of a social network
to any other via only a small number of intermediate acquaintances, the
exact number typically scaling as the logarithm of the total number of
individuals comprising the network.  Within the population of the world,
for example, it has been suggested that there are only about ``six degrees
of separation'' between any human being and any other\cite{Milgram67}.
This behavior is not seen in regular lattices but is a well-known property
of random graphs, where the average shortest path between two
randomly-chosen vertices scales as $\log N/\log z$, where $N$ is the total
number of vertices in the graph and $z$ is the average coordination
number\cite{Bollobas85}.

Recently, Watts and Strogatz\cite{WS98} have proposed a model which
attempts to mimic the properties of social networks.  This ``small-world''
model consists of a network of vertices whose topology is that of a regular
lattice, with the addition of a low density $\phi$ of connections between
randomly-chosen pairs of vertices\cite{note1}.  Watts and Strogatz showed
that graphs of this type can indeed possess well-defined locales in the
sense described above while at the same time possessing average
vertex--vertex distances which are comparable with those found on true
random graphs, even for quite small values of~$\phi$.

%No exact solution for the average properties of the Watts--Strogatz model
%has yet been found, but it is possible nonetheless to say many things about
%the way the model behaves.

In this paper we study in detail the behavior of the small-world model,
concentrating particularly on its scaling properties.  The outline of the
paper is as follows.  In Section~\ref{model} we define the model.  In
Section~\ref{length} we study the typical length-scales present in the
model and argue that the model undergoes a continuous phase transition as
the density of random connections tends to zero.  We also examine the
cross-over been large- and small-world behavior in the model, and the
structure of ``neighborhoods'' of adjacent vertices.  In
Section~\ref{scaling} we derive a scaling form for the average
vertex--vertex distance on a small-world graph and demonstrate numerically
that this form is followed over a wide range of the parameters of the
model.  In Section~\ref{seceffdim} we calculate the effective dimension of
small-world graphs and show that this dimension depends on the length-scale
on which we examine the graph.  In Section~\ref{percolation} we consider
the properties of site percolation on these systems, as a model of the
spread of information or disease through social networks.  Finally, in
Section~\ref{concs} we give our conclusions.

\section{The small-world model}
\label{model}
The original small-world model of Watts and Strogatz, in its simplest
incarnation, is defined as follows.  We take a one-dimensional lattice of
$L$ vertices with connections or bonds between nearest neighbors and
periodic boundary conditions (the lattice is a ring).  Then we go through
each of the bonds in turn and independently with some probability $\phi$
``rewire'' it.  Rewiring in this context means shifting one end of the bond
to a new vertex chosen uniformly at random from the whole lattice, with the
exception that no two vertices can have more than one bond running between
them, and no vertex can be connected by a bond to itself.  In this model
the average coordination number $z$ remains constant ($z=2$) during the
rewiring process, but the coordination number of any particular vertex may
change.  The total number of rewired bonds, which we will refer to as
``shortcuts'', is $\phi L$ on average.

For the purposes of analytic treatment the Watts--Strogatz model has a
number of problems.  One problem is that the distribution of shortcuts is
not completely uniform; not all choices of the positions of the rewired
bonds are equally probable.  For example, configurations with more than one
bond between a particular pair of vertices are explicitly forbidden.  This
non-uniformity of the distribution makes an average over different
realizations of the randomness hard to perform.

\begin{figure}
\begin{center}
\psfig{figure=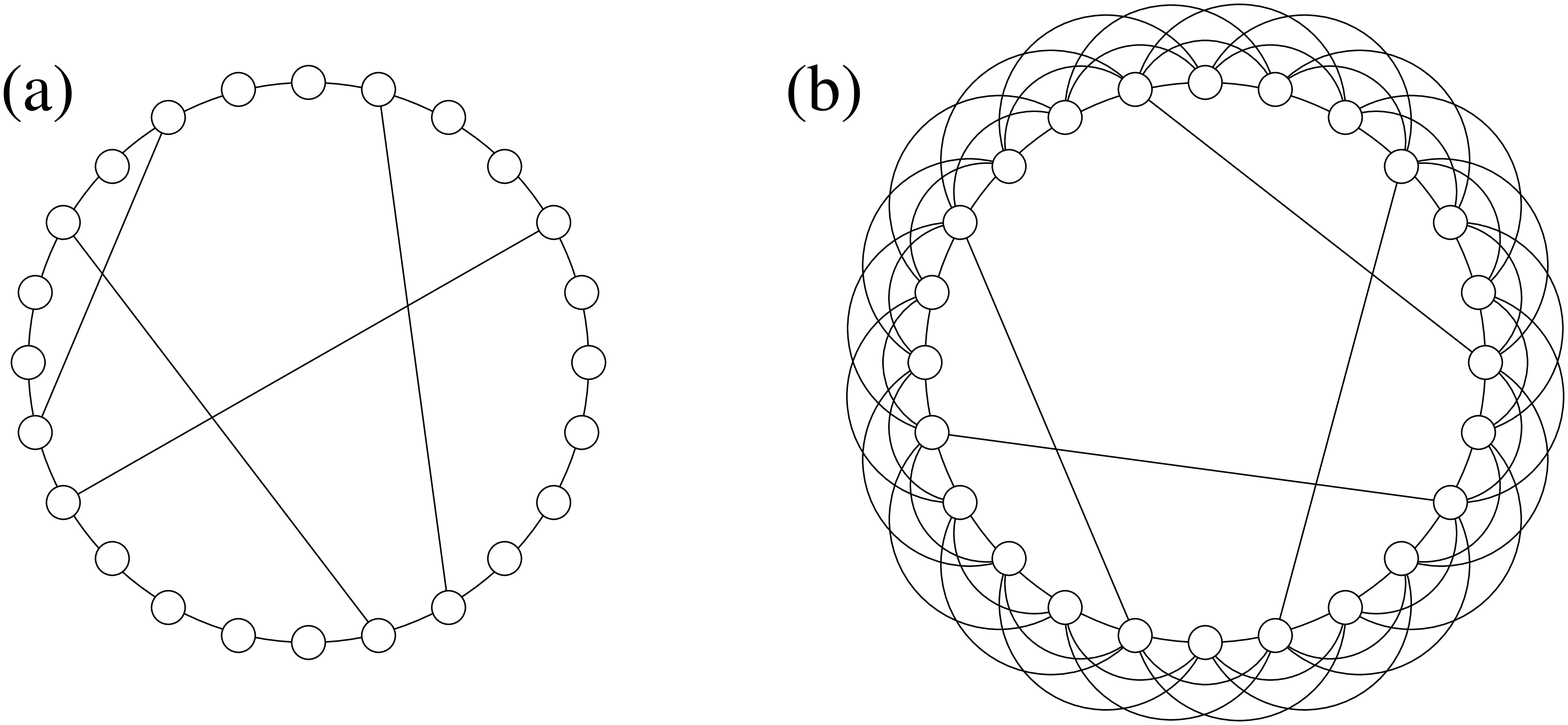,width=\columnwidth}
\end{center}
\caption{(a) An example of a small-world graph with $L=24$, $k=1$ and, in
  this case, four shortcuts.  (b)~An example with $k=3$.}
\label{onedim}
\end{figure}

A more serious problem is that one of the crucial quantities of interest in
the model, the average distance between pairs of vertices on the graph, is
poorly defined.  The reason is that there is a finite probability of a
portion of the lattice becoming detached from the rest in this model.
Formally, we can represent this by saying that the distance from such a
portion to a vertex elsewhere on the lattice is infinite.  However, this
means that the average vertex--vertex distance on the lattice is then
itself infinite, and hence that the vertex--vertex distance averaged over
all realizations is also infinite.  For numerical studies such as those of
Watts and Strogatz this does not present any substantial difficulties, but
for analytic work it results in a number of quantities and expressions
being poorly defined.

Both of these problems can be circumvented by a slight modification of the
model.  In our version of the small-world model we again start with a
regular one-dimensional lattice, but now instead of rewiring each bond with
probability $\phi$, we add shortcuts between pairs of vertices chosen
uniformly at random but we do not remove any bonds from the regular
lattice.  We also explicitly allow there to be more than one bond between
any two vertices, or a bond which connects a vertex to itself.  In order to
preserve compatibility with the results of Watts and Strogatz and others,
we add with probability $\phi$ one shortcut for each bond on the original
lattice, so that there are again $\phi L$ shortcuts on average.  The
average coordination number is $z=2(1+\phi)$.  This model is equivalent to
the Watts--Strogatz model for small $\phi$, whilst being better behaved
when $\phi$ becomes comparable to~1.  Fig.~\ref{onedim}(a) shows one
realization of our model for $L=24$.

Real social networks usually have average coordination numbers $z$
significantly higher than~$2$, and we can arrange for higher $z$ in our
model in a number of ways.  Watts and Strogatz\cite{WS98} proposed adding
bonds to next-nearest or further neighbors on the underlying
one-dimensional lattice up to some fixed range which we will call
$k$\cite{note2}.  In our variation on the model we can also start with such
a lattice and then add shortcuts to it.  The mean number of shortcuts is
then $\phi kL$ and the average coordination number is $z=2k(1+\phi)$.
Fig.~\ref{onedim}(b) shows a realization of this model for $k=3$.

\begin{figure}
\begin{center}
\psfig{figure=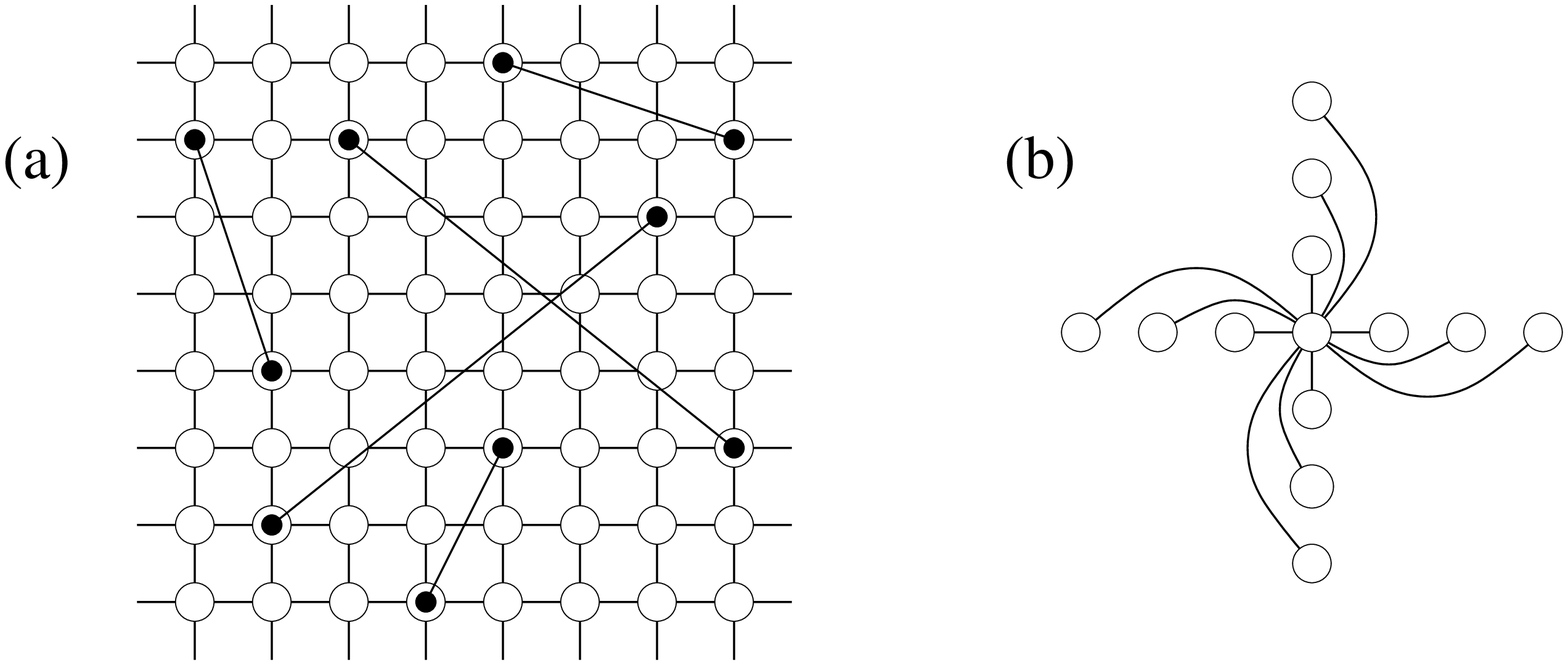,width=\columnwidth}
\end{center}
\caption{(a) An example of a $k=1$ small-world graph with an underlying
  lattice of dimension $d=2$.  (b)~The pattern of bonds around a vertex on
  the $d=2$ lattice for $k=3$.}
\label{twodim}
\end{figure}

Another way of increasing the coordination number, suggested first by
Watts\cite{WattsThesis,Watts99}, is to use an underlying lattice for the
model with dimension greater than one.  In this paper we will consider
networks based on square and (hyper)cubic lattices in $d$ dimensions.  We
take a lattice of linear dimension $L$, with $L^d$ vertices,
nearest-neighbor bonds and periodic boundary conditions, and add shortcuts
between randomly chosen pairs of vertices.  Such a graph has $\phi dL^d$
shortcuts and an average coordination number $z=2d(1+\phi)$.  An example is
shown in Fig.~\ref{twodim}(a) for $d=2$.  We can also add bonds between
next-nearest or further neighbors to such a lattice.  The most
straightforward generalization of the one-dimensional case is to add bonds
along the principal axes of the lattice up to some fixed range $k$, as
shown in Fig.~\ref{twodim}(b) for $k=3$.  Graphs of this type have $\phi
kdL^d$ shortcuts on average and a mean coordination number of
$z=2kd(1+\phi)$.

Our main interest in this paper is with the properties of the small-world
model for small values of the shortcut probability $\phi$.  Watts and
Strogatz\cite{WS98} found that the model displays many of the
characteristics of true random graphs even for $\phi\ll1$, and it seems to
be in this regime that the model's properties are most like those of
real-world social networks.

\section{Length-scales in small-world graphs}
\label{length}
A fundamental observable property of interest on small-world lattices is
the shortest path between two vertices---the number of degrees of
separation---measured as the number of bonds traversed to get from one
vertex to another, averaged over all pairs of vertices and over all
realizations of the randomness in the model.  We denote this quantity
$\ell$.  On ordinary regular lattices $\ell$ scales linearly with the
lattice size $L$.  On the underlying lattices used in the models described
here for instance, it is equal to $\frac14 dL/k$.  On true random graphs,
in which the probability of connection between any two vertices is the
same, $\ell$ is proportional to $\log N/\log z$, where $N$ is the number of
vertices on the graph\cite{Bollobas85}.  The small-world model interpolates
between these extremes, showing linear scaling $\ell\sim L$ for small
$\phi$, or on systems small enough that there are very few shortcuts, and
logarithmic scaling $\ell\sim\log N = d\log L$ when $\phi$ or $L$ is large
enough.  In this section and the following one we study the nature of the
cross-over between these two regimes, which we refer to as ``large-world''
and ``small-world'' regimes respectively.  For simplicity we will work
mostly with the case $k=1$, although we will quote results for $k>1$ where
they are of interest.

When $k=1$ the small-world model has only one independent parameter---the
probability $\phi$---and hence can have only one non-trivial length-scale
other than the lattice constant of the underlying lattice.  This
length-scale, which we will denote $\xi$, can be defined in a number of
different ways, all definitions being necessarily proportional to one
another.  One simple way is to define $\xi$ to be the typical distance
between the ends of shortcuts on the lattice.  In a one-dimensional system
with $k=1$, for example, there are on average $\phi L$ shortcuts and
therefore $2\phi L$ ends of shortcuts.  Since the lattice has $L$ vertices,
the average distance between ends of shortcuts is $L/(2\phi L) =
1/(2\phi)$.  In fact, it is more convenient for our purposes to define
$\xi$ without the factor of $2$ in the denominator, so that $\xi=1/\phi$,
or for general $d$
\begin{equation}
\xi = {1\over(\phi d)^{1/d}}.
\label{defsxi}
\end{equation}
For $k>1$ the appropriate generalization is\cite{note3}
\begin{equation}
\xi = {1\over (\phi kd)^{1/d}}.
\label{generalxi}
\end{equation}
As we see, $\xi$ diverges as $\phi\to0$ according to\cite{note4}
\begin{equation}
\xi \sim \phi^{-\tau},
\label{defstau}
\end{equation}
where the exponent $\tau$ is
\begin{equation}
\tau = {1\over d}.
\label{valuetau}
\end{equation}

A number of authors have previously considered a divergence of the kind
described by Eq.~\eref{defstau} with $\xi$ defined not as the typical
distance between the ends of shortcuts, but as the system size $L$ at which
the cross-over from large- to small-world scaling
occurs\cite{BA99,Barrat99,NW99,MMP99}.  We will shortly argue that in fact
the length-scale $\xi$ defined here is precisely equal to this cross-over
length, and hence that these two divergences are the same.

The quantity $\xi$ plays a role similar to that of the correlation length
in an interacting system in standard statistical physics.
%The correspondence
%is not complete and for this reason we will not actually refer to $\xi$ as
%the correlation length.  Nonetheless, the analogy is helpful.
Its leaves the system with no length-scale other than the lattice spacing,
so that at long distances we expect all spatial distributions to be
scale-free.  This is precisely the behavior one sees in an interacting
system undergoing a continuous phase transition, and it is reasonable to
regard the small-world model as having a continuous phase transition at
this point.  Note that the transition is a one-sided one since $\phi$ is a
probability and cannot take values less than zero.  In this respect the
transition is similar to that seen in the one-dimensional Ising model, or
in percolation on a one-dimensional lattice.  The exponent $\tau$ plays the
part of a critical exponent for the system, similar to the correlation
length exponent $\nu$ for a thermal phase transition.

De Menezes~\etal\cite{MMP99} have argued that the length-scale $\xi$ can
{\em only\/} be defined in terms of the cross-over point between large- and
small-world behavior, that there is no definition of $\xi$ which can be
made consistent in the limit of large system size.  For this reason they
argue that the transition at $\phi=0$ should be regarded as first-order
rather than continuous.  In fact however, the arguments of
de~Menezes~\etal\ show only that one particular definition of $\xi$ is
inconsistent; they show that $\xi$ cannot be consistently defined in terms
of the mean vertex--vertex distance between vertices in finite regions of
infinite small-world graphs.  This does not prove that no definition of
$\xi$ is consistent in the $L\to\infty$ limit and, as we have demonstrated
here, consistent definitions do exist.  Thus it seems appropriate to
consider the transition at $\phi=0$ to be a continuous one.

Barth\'el\'emy and Amaral\cite{BA99} have conjectured on the basis of
numerical simulations that $\tau=\frac23$ for $d=1$.  As we have shown
here, $\tau$ is in fact equal to $1/d$, and specifically $\tau=1$ in one
dimension.  We have also demonstrated this result previously using a
renormalization group~(RG) argument\cite{NW99}, and it has been confirmed
by extensive numerical simulations\cite{Barrat99,NW99,MMP99}.

The length-scale $\xi$ governs a number of other properties of small-world
graphs.  First, as mentioned above, it defines the point at which the
average vertex--vertex distance $\ell$ crosses over from linear to
logarithmic scaling with system size $L$.  This statement is necessarily
true, since $\xi$ is the only non-trivial length scale in the model, but we
can demonstrate it explicitly by noting that the linear scaling regime is
the one in which the average number of shortcuts on the lattice is small
compared with unity and the logarithmic regime is the one in which it is
large\cite{WattsThesis}.  The cross-over occurs in the region where the
average number of shortcuts is about one, or in other words when $\phi k
dL^d=1$.  Rearranging for $L$, the cross-over length is
\begin{equation}
L = {1\over(\phi k d)^{1/d}} = \xi.
\end{equation}

\begin{figure}
\begin{center}
\psfig{figure=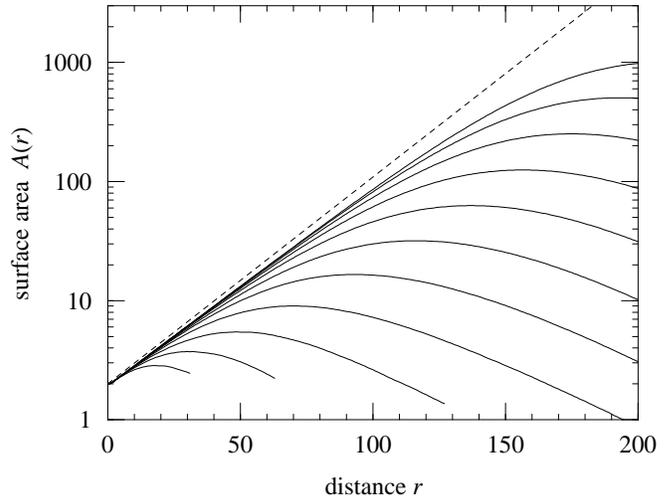,width=\columnwidth}
\end{center}
\caption{The mean surface area $A(r)$ of a neighborhood of radius $r$ on a
  $d=1$ small-world graph with $\phi=0.01$ for $L=128\ldots131\,072$ (solid
  lines).  The measurements are averaged over $1000$ realizations of the
  system each.  The dotted line is the theoretical result for $L=\infty$,
  Eq.~\eref{area}.}
\label{converge}
\end{figure}

\begin{figure}
\begin{center}
\psfig{figure=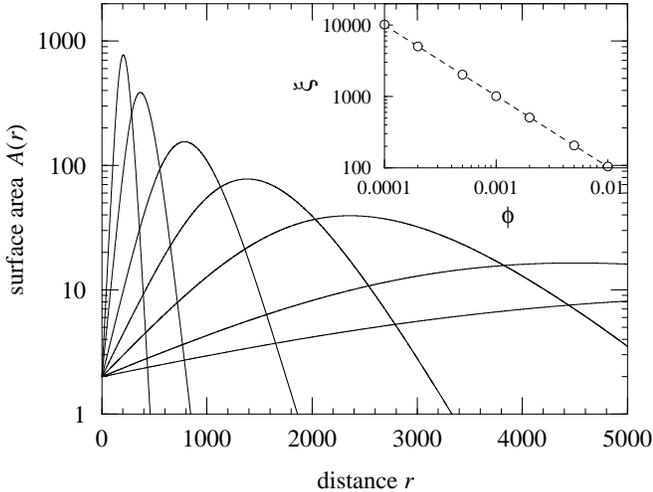,width=\columnwidth}
\end{center}
\caption{The mean surface area $A(r)$ of a neighborhood of radius $r$ on a
  $d=1$ small-world graph with $L=100\,000$ for
  $\phi=10^{-4}\ldots10^{-2}$.  The measurements are averaged over $1000$
  realizations of the system each.  Inset: the value of $\xi$ extracted
  from the curves in the main figure, as a function of $\phi$.  The
  gradient of the line gives the value of the exponent $\tau$, which is
  found by a least squares fit (the dotted line) to be $0.99\pm0.01$.}
\label{radii}
\end{figure}

The length-scale $\xi$ also governs the average number $V(r)$ of neighbors
of a given vertex within a neighborhood of radius $r$.  The number of
vertices in such a neighborhood increases as $r^d$ for $r\ll\xi$ while for
$r\gg\xi$ the graph behaves as a random graph and the size of the
neighborhood must increase exponentially with some power of $r/\xi$.  To
derive the specific functional form of $V(r)$ we consider a small-world
graph in the limit of infinite $L$.  Let $a(r)$ be the surface area of a
``sphere'' of radius $r$ on the underlying lattice of the model, i.e.,~it
is the number of points which are exactly $r$ steps away from any vertex.
(For $k=1$, $a(r) = 2^d r^{d-1}/\Gamma(d)$ when $r\gg1$.)  The volume
within a neighborhood of radius $r$ in an infinite system is the sum of
$a(r)$ over $r$, plus a contribution of $V(r-r')$ for every shortcut
encountered at a distance $r'$, of which there are on average $2\xi^{-d}
a(r')$.  Thus $V(r)$ is in general the solution of the equation
\begin{equation}
V(r) = \sum_{r'=0}^r a(r') [1+2\xi^{-d} V(r-r')].
\end{equation}
In one dimension with $k=1$, for example, $a(r)=2$ for all $r$ and,
approximating the sum with an integral and then differentiating with
respect to $r$, we get
\begin{equation}
{\d V\over\d r} = 2[1 + 2V(r)/\xi],
\end{equation}
which has the solution
\begin{equation}
V(r) = \half\xi(\e^{4r/\xi} - 1).
\label{volume}
\end{equation}
Note that for $r\ll\xi$ this scales as $r$, independent of $\xi$, and for
$r\gg\xi$ it grows exponentially, as expected.  Eq.~\eref{volume} also
implies that the surface area of a sphere of radius $r$ on the graph, which
is the derivative of $V(r)$, should be
\begin{equation}
A(r) = 2\e^{4r/\xi}.
\label{area}
\end{equation}
These results are easily checked numerically and give us a simple
independent measurement of $\xi$ which we can use to confirm our earlier
arguments.  In Fig.~\ref{converge} we show curves of $A(r)$ from computer
simulations of systems with $\phi=0.01$ for values of $L$ equal to powers
of two from $128$ up to $131\,072$ (solid lines).  The dotted line is
Eq.~\eref{area} with $\xi$ taken from Eq.~\eref{defsxi}.  The convergence
of the simulation results to the predicted exponential form as the system
size grows confirms our contention that $\xi$ is well-defined in the limit
of large $L$.  Fig.~\ref{radii} shows $A(r)$ for $L=100\,000$ for various
values of $\phi$.  Eq.~\eref{area} implies that the slope of the lines in
the limit of small $r$ is $4/\xi$.  In the inset we show the values of
$\xi$ extracted from fits to the slope as a function of $\phi$ on
logarithmic scales, and a straight-line fit to these points gives us an
estimate of $\tau=0.99\pm0.01$ for the exponent governing the transition at
$\phi=0$ (Eq.~\eref{defstau}).  This is in good agreement with our
theoretical prediction that $\tau=1$.

\section{Scaling in small-world graphs}
\label{scaling}
Given the existence of the single non-trivial length-scale $\xi$ for the
small-world model, we can also say how the mean vertex--vertex distance
$\ell$ should scale with system size and other parameters near the phase
transition.  In this regime the dimensionless quantity $\ell/L$ can be a
function only of the dimensionless quantity $L/\xi$, since no other
dimensionless combinations of variables exist.  Thus we can write
\begin{equation}
\ell = L f(L/\xi),
\end{equation}
where $f(x)$ is an unknown but universal scaling function.  A scaling form
similar to this was suggested previously by Barth\'el\'emy and
Amaral\cite{BA99} on empirical grounds.  Substituting from
Eq.~\eref{defsxi}, we then get for the $k=1$ case
\begin{equation}
\ell = L f(\phi^{1/d} L).
\end{equation}
(We have absorbed a factor of $d^{1/d}$ into the definition of $f(x)$ here
to make it consistent with the definition we used in
Ref.~\onlinecite{NW99}.)  The usefulness of this equation derives from the
fact that the function $f(x)$ contains no dependence on $\phi$ or $L$ other
than the explicit dependence introduced through its argument.  Its
functional form can however change with dimension $d$ and indeed it does.
In order to obey the known asymptotic forms of $\ell$ for large and small
systems, the scaling function $f(x)$ must satisfy
\begin{equation}
f(x) \sim {\log x\over x}\qquad\mbox{as $x\to\infty$},
\label{upperlim}
\end{equation}
and
\begin{equation}
f(x) \to \mbox{$\frac14$} d\qquad\mbox{as $x\to0$}.
\label{lowerlim}
\end{equation}

When $k>1$, $\ell$ tends to $\frac14 dL/k$ for small values of $L$ and
$\xi$ is given by Eq.~\eref{generalxi}, so the appropriate generalization
of the scaling form is
\begin{equation}
\ell = {L\over k} f\bigl((\phi k)^{1/d} L\bigr),
\label{scalinglaw}
\end{equation}
with $f(x)$ taking the same limiting forms~\eref{upperlim}
and~\eref{lowerlim}.  Previously we derived this scaling form in a more
rigorous way using an RG argument\cite{NW99}.

\begin{figure}
\begin{center}
\psfig{figure=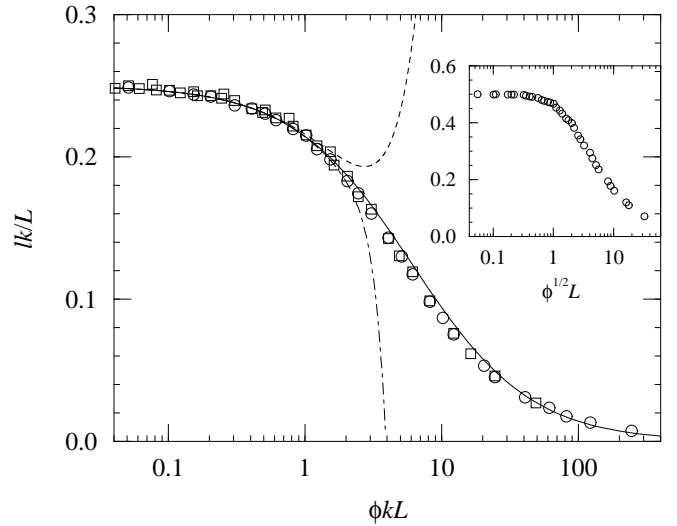,width=\columnwidth}
\end{center}
\caption{Data collapse for numerical measurements of the mean
  vertex--vertex distance on small-world graphs with $d=1$.  Circles and
  squares are results for $k=1$ and $k=5$ respectively for values of $L$
  between $128$ and $32\,768$ and values of $\phi$ between $1\times10^{-6}$
  and $3\times10^{-2}$.  Each point is averaged over 1000 realizations of
  the randomness.  In all cases the errors on the points are smaller than
  the points themselves.  The dashed line is the second-order series
  approximation with exact coefficients given in Eq.~\eref{series}, while
  the dot-dashed line is the fifth-order approximation using numerical
  results for the last three coefficients.  The solid line is the
  third-order Pad\'e approximant, Eqs.~\eref{pade} and~\eref{padesoln}.
  Inset: data collapse for two-dimensional systems with $k=1$ for values of
  $L$ from 64 to $1024$ and $\phi$ from $3\times10^{-6}$ up to
  $1\times10^{-3}$.}
\label{collapse}
\end{figure}

We can again test these results numerically by measuring $\ell$ on
small-world graphs for various values of $\phi$, $k$ and $L$.
Eq.~\eref{scalinglaw} implies that if we plot the results on a graph of
$\ell k/L$ against $(\phi k)^{1/d} L$, they should collapse onto a single
curve for any given dimension~$d$.  In Fig.~\ref{collapse} we have done
this for systems based on underlying lattices with $d=1$ for a range of
values of $\phi$ and $L$, for $k=1$ and~$5$.  As the figure shows, the
collapse is excellent.  In the inset we show results for $d=2$ with $k=1$,
which also collapse nicely onto a single curve.  The lower limits of the
scaling functions in each case are in good agreement with our theoretical
predictions of $\frac14$ for $d=1$ and $\frac12$ for $d=2$.

We are not able to solve exactly for the form of the scaling function
$f(x)$, but we can express it as a series expansion in powers of $\phi$ as
follows.  Since the scaling function is universal and has no implicit
dependence on $k$, it is adequate to calculate it for the case $k=1$; its
form is the same for all other values of $k$.  For $k=1$ the probability of
having exactly $m$ shortcuts on the graph is
\begin{equation}
P_m = \biggl({dL^d\atop m}\biggr) \,\phi^m (1-\phi)^{dL^d-m}.
\end{equation}
Let $\ell_m$ be the mean vertex--vertex distance on a graph with $m$
shortcuts in the limit of large $L$, averaged over all such graphs.  Then
the mean vertex--vertex distance averaged over all graphs regardless of the
number of shortcuts is
\begin{equation}
\ell = \sum_{m=0}^{dL^d} P_m \ell_m.
\label{expansion}
\end{equation}
Note that in order to calculate $\ell$ up to order $\phi^m$ we only need to
know the behavior of the model when it has $m$ or fewer shortcuts.  For the
$d=1$ case the values of the $\ell_m$ have been calculated up to $m=2$ by
Strang and Eriksson\cite{SE99} and are given in Table~\ref{table1}.
Substituting these into Eq.~\eref{expansion} and collecting terms in
$\phi$, we then find that
\begin{equation}
{\ell\over L} = \mbox{$\frac14 - \frac{1}{24} \phi L + \frac{11}{1440} \phi^2
  L^2 - \frac{11}{1440} \phi^2 L + \O(\phi^3)$}.
\end{equation}
The term in $\phi^2 L$ can be dropped when $L$ is large or $\phi$ small,
since it is negligible by comparison with at least one of the terms before
it.  Thus the scaling function is
\begin{equation}
f(x) = \mbox{$\frac14 - \frac{1}{24} x + \frac{11}{1440} x^2 + \O(x^3)$}.
\label{series}
\end{equation}
This form is shown as the dotted line in Fig.~\ref{collapse} and agrees
well with the numerical calculations for small values of the scaling
variable $x$, but deviates badly for large values.

\begin{table}
\begin{tabular}{cccc}
 & $m$ & $\ell_m/L$ & \\
\tableline
 & 0 & $1/4$ & \\
 & 1 & $5/24$ & \\
 & 2 & $131/720$ & \\
 & 3 & $0.1549\pm0.0003$ & \\
 & 4 & $0.1365\pm0.0003$ & \\
 & 5 & $0.1232\pm0.0003$ & \\
\end{tabular}
\null\vspace{3mm}
\caption{Average vertex--vertex distances per vertex $\ell_m/L$ on $d=1$
small-world graphs with exactly $m$ shortcuts and $k=1$.  Values up to
$m=2$ are the exact results of Strang and Eriksson\protect\cite{SE99}.
Values for $m=3\ldots5$ are our numerical results.}
\label{table1}
\end{table}

Calculating the exact values of the quantities $\ell_m$ for higher orders
is an arduous task and probably does not justify the effort involved.
However, we have calculated the values of the $\ell_m$ numerically up to
$m=5$ by evaluating the average vertex--vertex distance $\ell$ on graphs
which are constrained to have exactly 3, 4 or 5 shortcuts.  Performing a
Taylor expansion of $\ell/L$ about $L=\infty$, we get
\begin{equation}
{\ell\over L} = {\ell_m\over L}
                \Bigl[1 + {c\over L} + \O\bigl(L^{-2}\bigr)\Bigr],
\end{equation}
where $c$ is a constant.  Thus we can estimate $\ell_m/L$ from the
vertical-axis intercept of a plot of $\ell/L$ against $L^{-1}$ for large
$L$.  The results are shown in Table~\ref{table1}.  Calculating higher
orders still would be straightforward.

Using these values we have evaluated the scaling function $f(x)$ up to
fifth order in $x$; the result is shown as the dot--dashed line in
Fig.~\ref{collapse}.  As we can see the range over which it matches the
numerical results is greater than before, but not by much, indicating that
the series expansion converges only slowly as extra terms are added.  It
appears therefore that series expansion would be a poor way of calculating
$f(x)$ over the entire range of interest.

A much better result can be obtained by using our series expansion
coefficients to define a Pad\'e approximant to $f(x)$\cite{GG74,note6}.
Since we know that $f(x)$ tends to a constant $f(0)=\frac14 d$ for small
$x$ and falls off approximately as $1/x$ for large $x$, the appropriate
Pad\'e approximants to use are odd-order approximants where the approximant
of order $2n+1$ ($n$ integer) has the form
\begin{equation}
f(x) = f(0) {A_n(x)\over B_{n+1}(x)},
\end{equation}
where $A_n(x)$ and $B_n(x)$ are polynomials in $x$ of degree $n$ with
constant term equal to~1.  For example, to third order we should use the
approximant
\begin{equation}
f(x) = f(0) {1 + a_1 x\over1 + b_1 x + b_2 x^2}.
\label{pade}
\end{equation}
Expanding about $x=0$ this gives
\begin{eqnarray}
{f(x)\over f(0)} &=& 1 + (a_1 - b_1) x + (b_1^2 - a_1 b_1 - b_2) x^2\nonumber\\
  & & + [(a_1 - b_1) (b_1^2 - b_2) + b_1 b_2] x^3 + \O(x^4).\nonumber\\
\end{eqnarray}
Equating coefficients order by order in $x$ and solving for the $a$'s and
$b$'s, we find that
\begin{eqnarray}
a_1 &=& 1.825\pm0.075,\nonumber\\
\label{padesoln}
b_1 &=& 1.991\pm0.075,\\
b_2 &=& 0.301\pm0.012.\nonumber
\end{eqnarray}
Substituting these back into~\eref{pade} and using the known value of
$f(0)$ then gives us our approximation to $f(x)$.  This approximation is
plotted as the solid line in Fig.~\ref{collapse} and, as the figure shows,
is an excellent guide to the value of $f(x)$ over a large range of $x$.  In
theory it should be possible to calculate the fifth-order Pad\'e
approximant using the numerical results in Table~\ref{table1}, although we
have not done this here.  Substituting $f(x)$ back into the scaling form,
Eq.~\eref{scalinglaw}, we can also use the Pad\'e approximant to predict
the value of the mean vertex--vertex distance for any values of $\phi$, $k$
and $L$ within the scaling regime.  We will make use of this result in the
next section to calculate the effective dimension of small-world graphs.

\section{Effective dimension}
\label{seceffdim}
The calculation of the volumes and surface areas of neighborhoods of
vertices on small-world graphs in Section~\ref{length} leads us naturally
to the consideration of the dimension of these systems.  On a regular
lattice of dimension $D$, the volume $V(r)$ of a neighborhood of radius $r$
increases in proportion to $r^D$, and hence one can calculate $D$
from\cite{note7}
\begin{equation}
D = {\d\log V\over\d\log r} = {r A(r)\over V(r)},
\label{defsd1}
\end{equation}
where $A(r)$ is the surface area of the neighborhood, as previously.  We
can use the same expression to calculate the effective dimension of our
small-world graphs.  Thus in the case of an underlying lattice of dimension
$d=1$, the effective dimension of the graph is
\begin{equation}
D = {4r\over\xi}\,{\e^{4r/\xi}\over\e^{4r/\xi}-1},
\label{effdim}
\end{equation}
where we have made use of Eqs.~\eref{volume} and~\eref{area}.  For
$r\ll\xi$ this tends to one, as we would expect, and for $r\gg\xi$ it tends
to $4r/\xi$, increasing linearly with the radius of the neighborhood.  Thus
the effective dimension of a small-world graph depends on the length-scale
on which we look at it, in a way reminiscent of the behavior of
multifractals\cite{Mandelbrot74,HJKPS86}.  This result will become
important in Section~\ref{percolation} when we consider site percolation on
small-world graphs.

\begin{figure}
\begin{center}
\psfig{figure=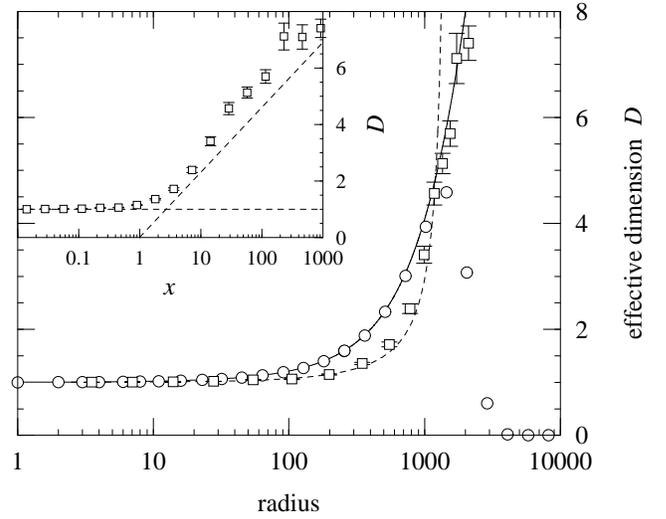,width=\columnwidth}
\end{center}
\caption{Effective dimension $D$ of small-world graphs.  The circles are
  results for $D$ from numerical calculations on an $L=1\,000\,000$ system
  with $d=1$, $k=1$ and $\phi=10^{-3}$ using Eq.~\eref{defsd1}.  The errors
  on the points are in all cases smaller than the points themselves.  The
  solid line is Eq.~\eref{effdim}.  The squares are calculated from
  Eq.~\eref{otherd} by numerical differentiation of simulation results for
  the scaling function $f(x)$ of one-dimensional systems.  The dotted line
  is Eq.~\eref{otherd} evaluated using the third-order Pad\'e approximant
  to the scaling function derived in Section~\ref{scaling}.  Inset:
  effective dimension from Eq.~\eref{otherd} plotted as a function of the
  scaling variable $x$.  The dotted lines represent the asymptotic forms
  for large and small $x$ discussed in the text.}
\label{dimension}
\end{figure}

In Fig.~\ref{dimension} we show the effective dimension of neighborhoods on
a large graph measured in numerical simulations (circles), along with the
analytic result, Eq.~\eref{effdim} (solid line).  As we can see from the
figure, the numerical and analytic results are in good agreement for small
radii $r$, but the numerical results fall off sharply for larger $r$.  The
reason for this is that Eq.~\eref{defsd1} breaks down as $V(r)$ approaches
the volume of the entire system; $V(r)$ must tend to $L^d$ in this limit
and hence the derivative in~\eref{defsd1} tends to zero.  The same effect
is also seen if one tries to use Eq.~\eref{defsd1} on ordinary regular
lattices of finite size.  To characterize the dimension of an entire system
therefore, we use another measure of $D$ as follows.

On a regular lattice of finite linear size $\ell$, the number of vertices
$N$ scales as $\ell^D$ and hence we can calculate the dimension from
\begin{equation}
D = {\d\log N\over\d\log\ell}.
\label{defsd2}
\end{equation}
We can apply the same formula to the calculation of the effective dimension
of small-world graphs putting $N=L^d$, although, since we don't have an
analytic solution for $\ell$, we cannot derive an analytic solution for $D$
in this case.  On the other hand, if we are in the scaling regime described
in Section~\ref{scaling}---the regime in which $\xi\gg1$---then
Eq.~\eref{scalinglaw} applies, along with the limiting forms,
Eqs.~\eref{upperlim} and~\eref{lowerlim}.  Substituting into~\eref{defsd2},
this gives us
\begin{equation}
{1\over D} = {\d\log\ell\over\d\log L^d}
           = {1\over d} \biggl[1+{\d\log f(x)\over\d\log x}\biggr],
\label{otherd}
\end{equation}
where $x = (\phi k)^{1/d} L \propto L/\xi$.  In other words $D$ is a
universal function of the scaling variable~$x$.  We know that $f(x)$ tends
to a constant for small $x$ (i.e.,~$\xi\gg L$), so that $D=d$ in this
limit, as we would expect.  For large $x$ (i.e.,~$\xi\ll L$),
Eq.~\eref{upperlim} applies.  Substituting into~\eref{otherd} this gives us
$D = d\log x$.  In the inset of Fig.~\ref{dimension} we show $D$ from
numerical calculations as a function of $x$ in one-dimensional systems of a
variety of sizes, along with the expected asymptotic forms, which it
follows reasonably closely.  In the main figure we also show this second
measure of $D$ (squares with error bars) as a function of the system radius
$\ell$ (with which it should scale linearly for large $\ell$, since
$\ell\sim\log x$ for large $x$).  As the figure shows, the two measures of
effective dimension agree reasonably well.  The numerical errors on the
first measure, Eq.~\eref{defsd1} are much smaller than those on the second,
Eq.~\eref{defsd2} (which is quite hard to calculate numerically), but the
second measure is clearly preferable as a measure of the dimension of the
entire system, since the first fails badly when $r$ approaches $\ell$.  We
also show the value of our second measure of dimension calculated using the
Pad\'e approximant to $f(x)$ derived in Section~\ref{scaling} (dotted line
in the main figure).  This agrees well with the numerical evaluation for
radii up to about $1000$ and has significantly smaller statistical error,
but overestimates $D$ somewhat beyond this point because of inaccuracies in
the approximation; the Pad\'e approximant scales as $1/x$ for large values
of $x$ rather than $\log x/x$, which means that $D$ will scale as $x$
rather than $\log x$ for large $x$.

\section{Percolation}
\label{percolation}
In the previous sections of this paper we have examined statistical
properties of small-world graphs such as typical length-scales,
vertex--vertex distances, scaling of volumes and areas, and effective
dimension of graphs.  These are essentially static properties of the
networks; to the extent that small-world graphs mimic social networks,
these properties tell us about the static structure of those networks.
However, social science also deals with dynamic processes going on within
social networks, such as the spread of ideas, information, or diseases.
This leads us to the consideration of dynamical models defined on
small-world graphs.  A small amount of research has already been conducted
in this area.  Watts\cite{WattsThesis,Watts99}, for instance, has
considered the properties of a number of simple dynamical systems defined
on small-world graphs, such as networks of coupled oscillators and cellular
automata.  Barrat and Weigt\cite{BW99} have looked at the properties of the
Ising model on small-world graphs and derived a solution for its partition
function using the replica trick.  Monasson\cite{Monasson99} looked at the
spectral properties of the Laplacian operator on small-world graphs, which
tells us about the time evolution of a diffusive field on the graph.
%(One
%can certainly imagine that some types of information might propagate across
%social networks in a diffusive fashion.)
There is also a moderate body of work in the mathematical and social
sciences which, although not directly addressing the small-world model,
deals with general issues of information propagation in networks, such as
the adoption of innovations\cite{Rogers62,CKM66,Strang91,Valente96}, human
epidemiology\cite{SS88,KM88,Logini88}, and the flow of data on the
Internet\cite{KW91,HK99}.

In this section we discuss the modeling of information or disease
propagation specifically on small-world graphs.  Suppose for example that
the vertices of a small-world graph represent individuals and the bonds
between them represent physical contact by which a disease can be spread.
The spread of ideas can be similarly modeled; the bonds then represent
information connections between individuals which could include letters,
telephone calls, or email, as well as physical contacts.  The simplest
model for the spread of disease is to have the disease spread between
neighbors on the graph at a uniform rate, starting from some initial
carrier individual.  From the results of Section~\ref{scaling} we already
know what this will look like.  If for example we wish to know how many
people in total have contracted a disease, that number is just equal to the
number $V(r)$ within some radius $r$ of the initial carrier, where $r$
increases linearly with time.  (We assume that no individual can catch the
disease twice, which is the case with most common diseases.)  Thus,
Eq.~\eref{volume} tells us that, for a $d=1$ small-world graph, the number
of individuals who have had a particular disease increases exponentially,
with a time-constant governed by the typical length-scale $\xi$ of the
graph.  Since all real-world social networks have a finite number of
vertices $N$, this exponential growth is expected to saturate when $V(r)$
reaches $N=L^d$.  This is not a particularly startling result; the usual
model for the spread of epidemics is the logistic growth model, which shows
initial exponential spread followed by saturation.

For a disease like influenza, which spreads fast but is self-limiting, the
number of people who are ill at any one time should be roughly proportional
to the area $A(r)$ of the neighborhood surrounding the initial carrier,
with $r$ again increasing linearly in time.  This implies that the epidemic
should have a single humped form with time, like the curves of $A(r)$
plotted in Fig.~\ref{radii}.  Note that the vertical axis in this figure is
logarithmic; on linear axes the curves are bell-shaped rather than
quadratic.  In the context of the spread of information or ideas, similar
behavior might be seen in the development of fads.  By a fad we mean an
idea which is catchy and therefore spreads fast, but which people tire of
quickly.  Fashions, jokes, toys, or buzzwords might be expected to show
popularity profiles over time similar to the curves in Fig.~\ref{radii}.

However, for most real diseases (or fads) this is not a very good model of
how they spread.  For real diseases it is commonly the case that only a
certain fraction $p$ of the population is susceptible to the disease.  This
can be mimicked in our model by placing a two-state variable on each vertex
which denotes whether the individual at that vertex is susceptible.  The
disease then spreads only within the local ``cluster'' of connected
susceptible vertices surrounding the initial carrier.  One question which
we can answer with such a model is how high the density $p$ of susceptible
individuals can be before the largest connected cluster covers a
significant fraction of the entire network and an epidemic ensues.

%In terms of information propagation, an equivalent problem is the
%following.  Suppose that a fraction $p$ of the population has access to a
%certain form of communication---the telephone for instance or, to take a
%contemporary example, electronic mail.  In the early eighties the fraction
%of people with email was very small, and even if one had an email account,
%the chances that one's friends had one too was rather small and
%communication possibilities were limited.  As we write this paper, on the
%eve of the third millennium however, it is clear that this has changed.
%The network has undergone a transition, and there are now sufficiently many
%people with access to email that it has become a common form of
%communication.

Mathematically, this is precisely the problem of site percolation on a
social network, at least in the case where the susceptible individuals are
randomly distributed over the vertices.  To the extent that small-world
graphs mimic social networks, therefore, it is interesting to look at the
percolation problem.  The transition corresponds to the point on a regular
lattice at which a percolating cluster forms whose size increases with the
size $L$ of the lattice for arbitrarily large $L$\cite{SA92}.  On random
graphs there is a similar transition, marked by the formation of a
so-called ``giant component'' of connected vertices\cite{AS92}.  On
small-world graphs we can calculate approximately the percolation
probability $p=p_c$ at which the transition takes place as follows.

Consider a $d=1$ small-world graph of the kind pictured in
Fig.~\ref{onedim}.  For the moment let us ignore the shortcut bonds and
consider the percolation properties just of the underlying regular lattice.
If we color in a fraction $p$ of the sites on this underlying lattice, the
occupied sites will form a number of connected clusters.  In order for two
adjacent parts of the lattice not to be connected, we must have a series of
at least $k$ consecutive unoccupied sites between them.  The number $n$ of
such series can be calculated as follows.  The probability that we have a
series of $k$ unoccupied sites starting at a particular site, followed by
an occupied one is $p(1-p)^k$.  Once we have such a series, the states of
the next $k$ sites are fixed and so it is not possible to have another such
series for $k$ steps.  Thus the number $n$ is given by
\begin{equation}
n = p(1-p)^k (L-kn).
\label{basicn}
\end{equation}
Rearranging for $n$ we get
\begin{equation}
n = L\,{p(1-p)^k\over1+kp(1-p)^k}.
\label{defsn}
\end{equation}
For this one-dimensional system, the percolation transition occurs when we
have just one break in the chain, i.e.,~when $n=1$.  This gives us a $k$th
order equation for $p_c$ which is in general not exactly soluble, but we
can find its roots numerically if we wish.

Now consider what happens when we introduce shortcuts into the graph.  The
number of breaks $n$, Eq.~\eref{defsn}, is also the number of connected
clusters of occupied sites on the underlying lattice.  Let us for the
moment suppose that the size of each cluster can be approximated by the
average cluster size.  A number $\phi kL$ of shortcuts are now added to the
graph between pairs of vertices chosen uniformly at random.  A fraction
$p^2$ of these will connect two occupied sites and therefore can connect
together two clusters of occupied sites.  The problem of when the
percolation transition occurs is then precisely that of the formation of a
giant component on an ordinary random graph with $n$ vertices.  It is known
that such a component forms when the mean coordination number of the random
graph is one\cite{AS92}, or alternatively, when the number of bonds on the
graph is a half the number of vertices.  In other words, the transition
probability $p_c$ must satisfy
\begin{equation}
p_c^2\phi kL = \half L\,{p_c(1-p_c)^k\over1+kp_c(1-p_c)^k},
\end{equation}
or
\begin{equation}
\phi = {(1-p_c)^k\over 2kp_c[1+kp_c(1-p_c)^k]}.
\label{solnpc}
\end{equation}

We have checked this result against numerical calculations.  In order to
find the value of $p_c$ numerically, we employ a tree-based invasion
algorithm similar to the invaded cluster algorithm used to find the
percolation point in Ising systems\cite{MCLSC95,BN99}.  This algorithm can
calculate the entire curve of average cluster size versus $p$ in time which
scales as $L\log L$\cite{note8}.  We define $p_c$ to be the point at which
the average cluster size divided by $L$ rises above a certain threshold.
For systems of infinite size the transition is instantaneous and hence the
choice of threshold makes no difference to $p_c$, except that $p_c$ can
never take a value lower than the threshold itself, since even in a fully
connected graph the average cluster size per vertex can be no greater than
the fraction $p_c$ of occupied vertices.  Thus it makes sense to choose the
threshold as low as possible.  In real calculations, however, we cannot use
an infinitesimal threshold because of finite size effects.  For the systems
studied here we have found that a threshold of $0.2$ works well.

\begin{figure}
\begin{center}
\psfig{figure=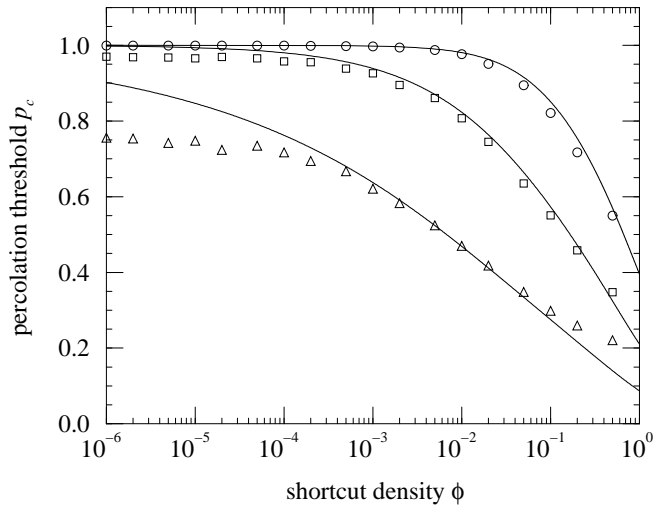,width=\columnwidth}
\end{center}
\caption{Numerical results for the percolation threshold on $L=10\,000$
  small-world graphs with $k=1$ (circles), 2~(squares), and 5~(triangles)
  as a function of the shortcut density $\phi$.  The solid lines are the
  analytic approximation to the same quantity, Eq.~\eref{solnpc}.}
\label{pc}
\end{figure}

Fig.~\ref{pc} shows the critical probability $p_c$ for systems of size
$L=10\,000$ for a range of values of $\phi$ for $k=1$, 2 and 5.  The points
are the numerical results and the solid lines are Eq.~\eref{solnpc}.  As
the figure shows the agreement between simulation and theory is good
although there are some differences.  As $\phi$ approaches one and the
value of $p_c$ drops, the two fail to agree because, as mentioned above,
$p_c$ cannot take a value lower than the threshold used in its calculation,
which was $0.2$ in this case.  The results also fail to agree for very low
values of $\phi$ where $p_c$ becomes large.  This is because
Eq.~\eref{defsn} is not a correct expression for the number of clusters on
the underlying lattice when $n<1$.  This is clear since when there are no
breaks in the sequence of connected vertices around the ring it is not also
true that there are no connected clusters.  In fact there is still one
cluster; the equality between number of breaks and number of clusters
breaks down at $n=1$.  The value of $p$ at which this happens is given by
putting $n=1$ in Eq.~\eref{basicn}.  Since $p$ is close to one at this
point its value is well approximated by
\begin{equation}
p \simeq 1 - L^{-1/k},
\end{equation}
and this is the value at which the curves in Fig.~\ref{pc} should roll off
at low $\phi$.  For $k=5$ for example, for which the roll-off is most
pronounced, this expression gives a value of $p\simeq0.8$, which agrees
reasonably well with what we see in the figure.

There is also an overall tendency in Fig.~\ref{pc} for our analytic
expression to overestimate the value of $p_c$ slightly.  This we put down
to the approximation we made in the derivation of Eq.~\eref{solnpc} that
all clusters of vertices on the underlying lattice can be assumed to have
the size of the average cluster.  In actual fact, some clusters will be
smaller than the average and some larger.  Since the shortcuts will connect
to clusters with probability proportional to the cluster size, we can
expect percolation to set in within the subset of larger-than-average
clusters before it would set in if all clusters had the average size.  This
makes the true value of $p_c$ slightly lower than that given by
Eq.~\eref{solnpc}.  In general however, the equation gives a good guide to
the behavior of the system.

We have also examined numerically the behavior of the mean cluster radius
$\rho$ for percolation on small-world graphs.  The radius of a cluster is
defined as the average distance between vertices within the cluster, along
the edges of the graph within the cluster.  This quantity is small for
small values of the percolation probability $p$ and increases with $p$ as
the clusters grow larger.  When we reach percolation and a giant component
forms it reaches a maximum value and then drops as $p$ increases further.
The drop happens because the percolating cluster is most filamentary when
percolation has only just set in and so paths between vertices are at their
longest.  With further increases in $p$ the cluster becomes more highly
connected and the average shortest path between two vertices decreases.

\begin{figure}
\begin{center}
\psfig{figure=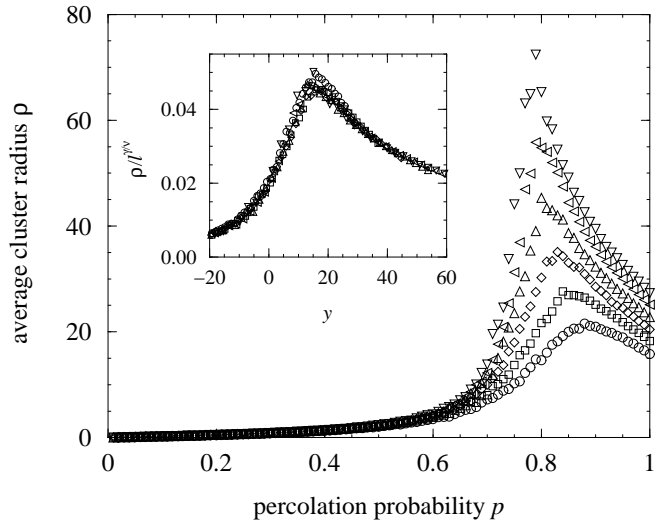,width=\columnwidth}
\end{center}
\caption{Average cluster radius $\rho$ as a function of the percolation
  probability $p$ for site percolation on small-world graphs with $k=1$,
  $\phi=0.1$ and $L$ equal to a power from $512$ up to $16\,384$ (circles,
  squares, diamonds, upward-pointing triangles, left-pointing triangles and
  downward-pointing triangles respectively).  Each set of points is
  averaged over 100 realizations of the corresponding graph.  Inset: the
  same data collapsed according to Eq.~\eref{rhoscale} with $\nu=0.59$,
  $\gamma=1.3$ and $p_c=0.74$.}
\label{rho}
\end{figure}

By analogy with percolation on regular lattices we might expect the average
cluster radius for a given value of $\phi$ to satisfy the scaling
form\cite{SA92}
\begin{equation}
\rho = \ell^{\gamma/\nu} \widetilde{\rho}\bigl((p-p_c) \ell^{1/\nu}\bigr),
\label{rhoscale}
\end{equation}
where $\widetilde{\rho}(x)$ is a universal scaling function, $\ell$ is the
radius of the entire system and $\gamma$ and $\nu$ are critical exponents.
In fact this scaling form is not precisely obeyed by the current system
because the exponents $\nu$ and $\gamma$ depend in general on the dimension
of the lattice.  As we showed in Section~\ref{seceffdim}, the dimension $D$
of a small-world graph depends on the length-scale on which you look at it.
Thus the value of $D$ ``felt'' by a cluster of radius $\rho$ will vary with
$\rho$, implying that $\nu$ and $\gamma$ will vary both with the
percolation probability and with the system size.  If we restrict ourselves
to a region sufficiently close to the percolation threshold, and to a
sufficiently small range of values of $\ell$, then Eq.~\eref{rhoscale}
should be approximately correct.

\begin{figure}
\begin{center}
\psfig{figure=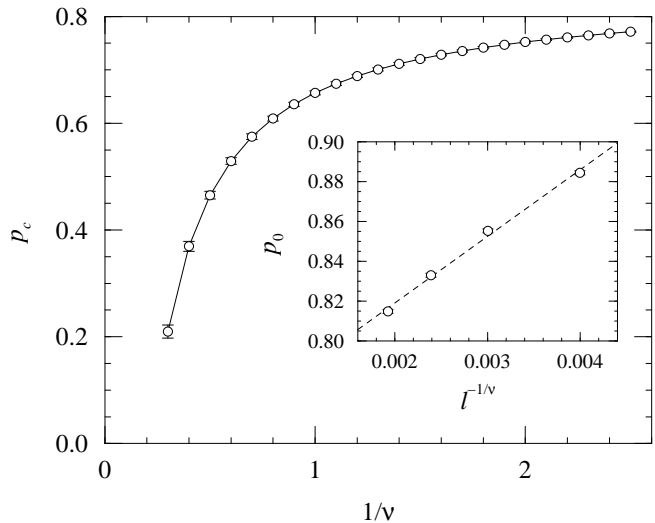,width=\columnwidth}
\end{center}
\caption{Best fit values of $p_c$ as a function of $1/\nu$.  Inset: the
  values are calculated from the vertical-axis intercept of a plot of the
  position $p_0$ of the peak of $\rho$ against $\ell^{-1/\nu}$ (see
  Eq.~\eref{intercept}).}
\label{nupc}
\end{figure}

In Fig.~\ref{rho} we show numerical data for $\rho$ for small-world graphs
with $k=1$, $\phi=0.1$ and $L$ equal to a power of two from 512 up to
$16\,384$.  As we can see, the data show the expected peaked form, with the
peak in the region of $p=0.8$, close to the expected position of the
percolation transition.  In order to perform a scaling collapse of these
data we need first to extract a suitable value of $p_c$.  We can do this by
performing a fit to the positions of the peaks in $\rho$\cite{NB99}.  Since
the scaling function $\widetilde{\rho}(x)$ is (approximately) universal,
the positions of these peaks all occur at the same value of the scaling
variable $y = (p-p_c) \ell^{1/\nu}$.  Calling this value $y_0$ and the
corresponding percolation probability $p_0$, we can rearrange for $p_0$ as
a function of $\ell$ to get
\begin{equation}
p_0 = p_c + y_0 \ell^{-1/\nu}.
\label{intercept}
\end{equation}
Thus if we plot the measured positions $p_0$ as a function of
$\ell^{-1/\nu}$, the vertical-axis intercept should give us the
corresponding value of $p_c$.  We have done this for a single value of
$\nu$ in the inset to Fig.~\ref{nupc}, and in the main figure we show the
resulting values of $p_c$ as a function of $1/\nu$.  If we now perform our
scaling collapse, with the restriction that the values of $\nu$ and $p_c$
fall on this line, then the best coincidence of the curves for $\rho$ is
obtained when $p_c=0.74$ and $\nu=0.59\pm0.05$---see the inset to
Fig.~\ref{rho}.  The value of $\gamma$ can be found separately by requiring
the heights of the peaks to match up, which gives $\gamma=1.3\pm0.1$.  The
collapse is noticeably poorer when we include systems of size smaller than
$L=512$, and we attribute this not merely to finite size corrections to the
scaling form, but also to variation in the values of the exponents $\gamma$
and $\nu$ with the effective dimension of the percolating cluster.

The value $p_c=0.74$ is in respectable agreement with the value of $0.82$
from our direct numerical measurements.  We note that $\nu$ is expected to
tend to $\half$ in the limit of an infinite-dimensional system.  The value
$\nu=0.59$ found here therefore confirms our contention that small-world
graphs have a high effective dimension even for quite moderate values of
$\phi$, and thus are in some sense close to being random graphs.  (On a
two-dimensional lattice by contrast $\nu=\frac43$.)

%This result is in accord with the results of Watts
%and Strogatz\cite{WS98}, who found that small-world graphs display many of
%the properties of random graphs even for relatively small values of $\phi$.

\section{Conclusions}
\label{concs}
In this paper we have studied the small-world network model of Watts and
Strogatz, which mimics the behavior of networks of social interactions.
Small-world graphs consist of a set of vertices joined together in a
regular lattice, plus a low density of ``shortcuts'' which link together
pairs of vertices chosen at random.  We have looked at the scaling
properties of small-world graphs and argued that there is only one typical
length-scale present other than the fundamental lattice constant, which we
denote $\xi$ and which is roughly the typical distance between the ends of
shortcuts.  We have shown that this length-scale governs the transition of
the average vertex--vertex distance on a graph from linear to logarithmic
scaling with increasing system size, as well as the rate of growth of the
number of vertices in a neighborhood of fixed radius about a given point.
We have also shown that the value of $\xi$ diverges on an infinite lattice
as the density of shortcuts tends to zero, and therefore that the system
possesses a continuous phase transition in this limit.  Close to the phase
transition, where $\xi$ is large, we have shown that the average
vertex--vertex distance on a finite graph obeys a simple scaling form and
in any given dimension is a universal function of a single scaling variable
which depends on the density of shortcuts, the system size and the average
coordination number of the graph.  We have calculated the form of the
scaling function to fifth order in the shortcut density using a series
expansion and to third order using a Pad\'e approximant.  We have defined
two measures of the effective dimension $D$ of small-world graphs and find
that the value of $D$ depends on the scale on which you look at the graph
in a manner reminiscent of the behavior of multifractals.  Specifically, at
length-scales shorter than $\xi$ the dimension of the graph is simply that
of the underlying lattice on which it is built, and for length-scales
larger than $\xi$ it increases linearly, with a characteristic constant
proportional to $\xi$.  The value of $D$ increases logarithmically with the
number of vertices in the graph.  We have checked all of these results by
extensive numerical simulation of the model and in all cases we find good
agreement between the analytic predictions and the simulation results.

In the last part of the paper we have looked at site percolation on
small-world graphs as a model of the spread of information or disease in
social networks.  We have derived an approximate analytic expression for
the percolation probability $p_c$ at which a ``giant component'' of
connected vertices forms on the graph and shown that this agrees well with
numerical simulations.  We have also performed extensive numerical
measurements of the typical radius of connected clusters on the graph as a
function of the percolation probability and shown by performing a scaling
collapse that these obey, to a reasonable approximation, the expected
scaling form in the vicinity of the percolation transition.  The
characteristic exponent $\nu$ takes a value close to $\half$, indicating
that, as far as percolation is concerned, the graph's properties are close
to those of a random graph.

\section*{Acknowledgments}
We thank Luis Amaral, Alain Barrat, Marc Barth\'el\'emy, Roman Koteck\'y,
Marcio de Menezes, Cris Moore, Cristian Moukarzel, Thadeu Penna, and Steve
Strogatz for helpful comments and conversations, and Gilbert Strang and
Henrik Eriksson for communicating to us some results from their forthcoming
paper.  This work was supported in part by the Santa Fe Institute and by
funding from the NSF (grant number PHY--9600400), the DOE (grant number
DE--FG03--94ER61951), and DARPA (grant number ONR N00014--95--1--0975).


\begin{references}
\bibitem{note1}
  In previous work the letter $p$ has been used to denote the density of
  random connections, rather than $\phi$.  We use $\phi$ here however, to
  avoid confusion with the percolation probability introduced in
  Section~\ref{percolation}, which is also conventionally denoted $p$.
\bibitem{note2}
  Watts and Strogatz used the letter $k$ to refer to the average
  coordination number, but we will find it convenient to distinguish
  between the coordination number, which we call $z$, and the range of the
  bonds.  In one dimension $z=2k$ and in general $z=2dk$ for networks based
  on $d$-dimensional lattices.
\bibitem{note3}
  The exact definition of $\xi$ depends on how you measure lengths in the
  model.  The definition given here is appropriate if $\xi$ is measured in
  terms of the lattice constant of the underlying lattice.  It would
  however be reasonable to measure it in terms of the number of bonds
  traversed between the ends of two shortcuts.  Since we are measuring
  lattice size $L$ in terms of the underlying lattice constant rather than
  number of bonds, the present definition is the more appropriate one in
  our case, but it would be perfectly consistent to define both $\xi$ and
  $L$ to be a factor of $k$ smaller; all the physical results would work
  out the same.
\bibitem{note4}
  In a system of finite size the average distance between the ends of two
  shortcuts cannot be larger than $\half L$, so we cannot observe this
  divergence once $\xi$ is larger than this.
\bibitem{note7}
  We use the capital letter $D$ to denote the dimension here, to
  distinguish it from the dimension $d$ of the underlying lattice defined
  in Section~\ref{model}.
\end{references}
\end{document}